\documentclass[12pt]{article}
\usepackage[usenames,dvips]{pstcol}
\usepackage{color}
\usepackage{amssymb}
\usepackage{epsfig}
\usepackage{footnote}
\usepackage{longtable}
\usepackage{verbatim}
\usepackage{array,multirow}
\usepackage{titlesec}
\usepackage{graphicx}
\def \PT {P$_T$~}
\def \Pt {P$_T$~}

\def\pbarp{{$\bar p$}p~}
\def\SpbarpS {S{$\bar{p}$}pS}
\begin{document}

\begin{center}
{\Large Pisa and the Collider Detector at Fermilab: a  History
of the Establishment of Precision Physics With a Calorimetric
Spectrometer at a Hadron Collider}\\

Henry Frisch \\
Enrico Fermi Institute and Physics Department \\
University of Chicago\\
\today
\end{center}

\begin{abstract}

This is a personal and admittedly US-centric attempt to
summarize the foundational impact of the Pisa CDF Group on the conceptual
design, construction, and early operation of the CDF Detector at
Fermilab.  I have tried to go back to original documents where
possible.

\end{abstract}

\tableofcontents*

\begin{figure}[h]
\centering
\includegraphics[angle=0,width=0.90\textwidth]{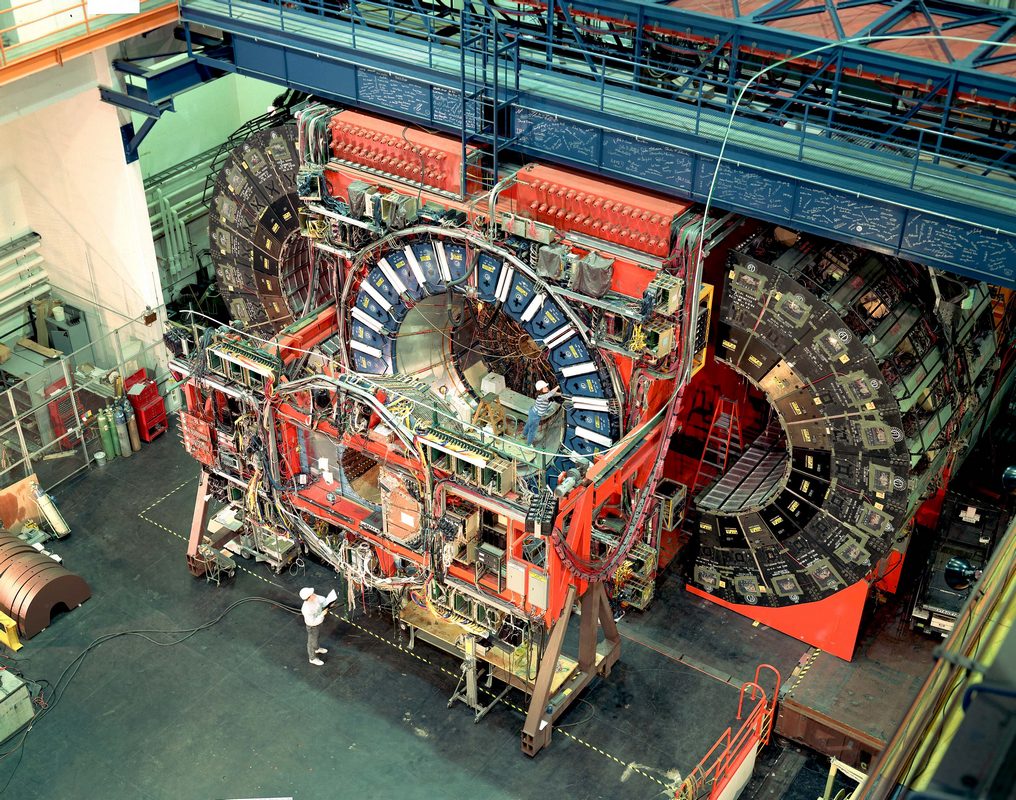}
\caption{The CDF Detector with the calorimeter arches open. }
\label{fig:cdf_arches}
\end{figure}

\section{The Impact of Pisa: Three Themes}
The formation and operation at Fermilab of the Collider Detector
Facility, as it was initially named spanned the development of the
modern picture of particle physics. Here I focus on three themes
to discuss the impact of the Pisa/Frascati group. Harder to
quantify, but most important, is the force of the senior Pisa
members towards the goal of a building an innovative world-class
detector in a  chaotic and unfunded context, described here at the
outset. The three themes of the talk are:

\begin{enumerate}
\setlength{\itemsep}{-0.05in}
\item The leadership role of Pisa in the innovative designs of the
   CDF tracking and calorimeter systems, and the subsequent construction.
\item The seminal development of precision mass measurements at hadron
  colliders using a magnetic spectrometer with precise tracking
  followed by projective calorimetry (aka `E/p', where `E' is the
  calorimeter energy measurement and `p' is the tracking momentum
  measurement).
\item The essential role of Pisa hardware in the discovery of the top
  quark and the extensive and path-breaking results on B-physics.
\end{enumerate}

In addition, Pisa played an essential role in leadership, management,
and drive. CDF was a wonderful experience, and among the many
wonderful groups, if I had to single out one that made the largest
contribution to the exceptionally high standards, creativity,
technical know-how, and collegiality of CDF, it would be Pisa.

\begin{figure}[h]
\centering
\includegraphics[angle=0,width=0.65\textwidth]{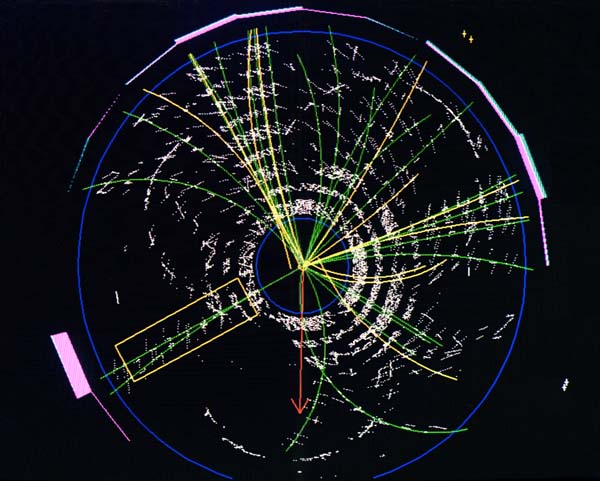}
\caption{A beams-eye view of a a pair of top quarks decaying into
  W-bosons and b-quark jets. The combination of a precise measurement
  of track momentum followed by a sign-independent measurement of
  the energy deposited pioneered high-precision measurements in
  complex events at hadron colliders.}
\label{fig:top_cot}
\end{figure}

Figure~\ref{fig:top_cot} shows a CDF event display of a beams-eye view
of a pair of top quarks decaying into W-bosons and b-quark jets. The
Pisa group made key contributions to the sophisticated trigger
system that selected the event for recording, to the calorimeters that
measured the (missing) transverse momentum carried off by the neutrino
in the leptonic decay of a W boson, and the silicon vertex detector
that allowed identification of the b-quarks, which are an essential
part of the signature of top quark production and decay.

\clearpage

\section{The Path to Precision Electroweak and Flavor Physics at Hadron Colliders}
The founding of
CDF was followed by a remarkable evolution in sophistication and precision
of detectors and techniques at the hadron colliders at Fermilab and
CERN since the pioneering experiments at the ISR and \SpbarpS.  Pisa
played a crucial leadership role in this evolution.

The following steps and the initial context provide an outline for my
talk. Understanding the intellectual and political context is
essential to appreciating the intellectual and collaborative role of
the Pisa/Frascatti group in the ultimate success of the Tevatron
Collider program.
\begin{enumerate}
\setlength{\itemsep}{-0.03in}
\item The start of high-\Pt physics (in the US): Bjorken, Feynman, Cronin
\item The chaotic road to the choice of \pbarp collisions in the Tevatron
\item Cronin starts the Collider Experiment Dept.: ZGS-MR (pp), MR
  (\pbarp), Tevatron (\pbarp)
\item Collider Detector: Giorgio Bellettini, Pisa MOUs
\item The development of precision measurements at a hadron
  collider: calorimeter behind a precision tracking
  system: the E/p method
\item The silicon Vertex Detector (SVX), Silicon Vertex Tracker (SVT)-
  real-time tracking, vertex-selected triggering
\item CDF (Pisa) footprint on hadron collider detector development
\item W and Z precision mass measurements; top quark discovery; B$_s$
  mixing, development of Higgs and BTSM search strategies
\end{enumerate}

\section{Hard Parton Scattering-- 1971: Bjorken, Feynman and Field, Cronin}

\begin{figure}[h]
\centering
\includegraphics[angle=0,width=1.0\textwidth]{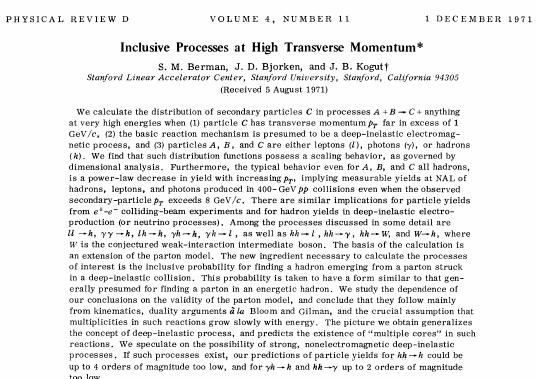}
\caption{The title page of the 1971 seminal paper by J. D.Bjorken,
S. Berman, and J. Kogut that lays out the landscape of hadron collider
physics: parton-parton scattering, Drell-Yan annihilation, and W/Z
boson production. This was the physics in which CDF was founded to explore
and discover.}
\label{fig:bbk_titlepage}
\end{figure}

The title page of a seminal paper from the summer of 1971 by Bjorken,
Berman, and Kogut that still reads remarkably well is shown in
Figure~\ref{fig:bbk_titlepage}. The title is `Inclusive Processes at
High Transverse Momentum'; the paper describes scattering, lepton pair
production, and W and Z production, and to a large extent still
describes the main thrust in HEP today.

\begin{figure}[t]
\centering
\includegraphics[angle=0,height=0.40\textheight]{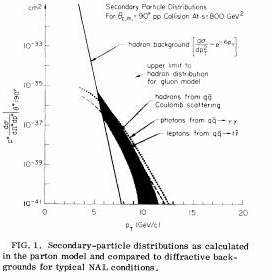}
\hfil
\includegraphics[angle=0,height=0.40\textheight]{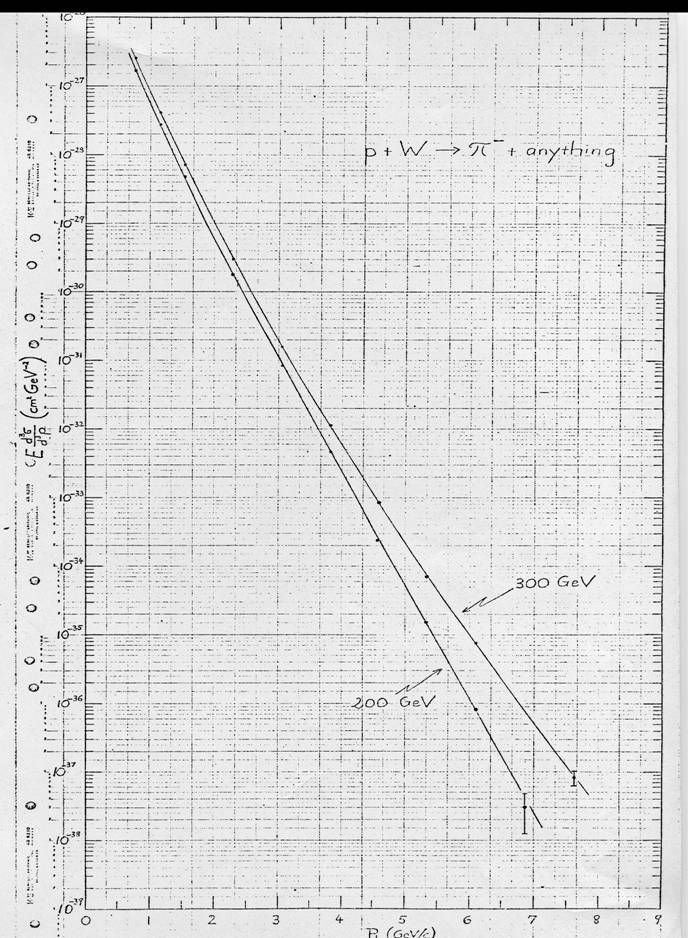}
\caption{Left: Predictions of high-\Pt particle spectra from
parton-parton scattering from the 1971 paper of Berman,
  Bjorken, and Kogut. The later designs of the CDF calorimeter tower geometry
  and calorimetric trigger were influenced by the single
  particle measurements at CERN and Fermilab as
 the properties of jets were still hotly debated at the time of the
 start of CDF. Right: The first measurements of inclusive hadron production at high
  \Pt at Fermilab by the Fermilab Experiment E100, with Jim Cronin and
Pierre Pirou\'{e} as spokesman. Cronin later become Head of the
Collider Detector Department, preceding the CDF Technical Design
Report and later proposal.} \label{fig:bbk_E100_pt_spectra}
\end{figure}

The left-hand panel of Figure~\ref{fig:bbk_E100_pt_spectra} shows
the prevailing view of particle production from that era, which
was an extrapolation from low $P_T$ of an exponential, typically
quoted as $e^{-6P_T}$, and their predicted power law deviations
above the prediction.

Presciently, Jim Cronin and Pierre Pirou\'{e} submitted a proposal to
Fermilab in 1970 to measure the spectrum of inclusive particle
production at high \Pt: electrons, muons, identified hadrons (pi, K,
and proton and their antiparticles), as well as short-lived new
particles. This was approved and became experiment E100, in the
Proton Lab.


The right-hand panel of Figure~\ref{fig:bbk_E100_pt_spectra} shows
the first results on inclusive hadron production at high \Pt at
Fermilab by E100, showing the clear deviation from the naive
exponential prediction, and, strikingly, the energy-dependence of
the cross-section at high \Pt.  However, due to delays in bringing
the Fermilab machine online, the CERN ISR collider beat us to the
punch, sadly, and barely.

\begin{figure}[h]
\centering
\includegraphics[angle=0,width=0.98\textwidth]{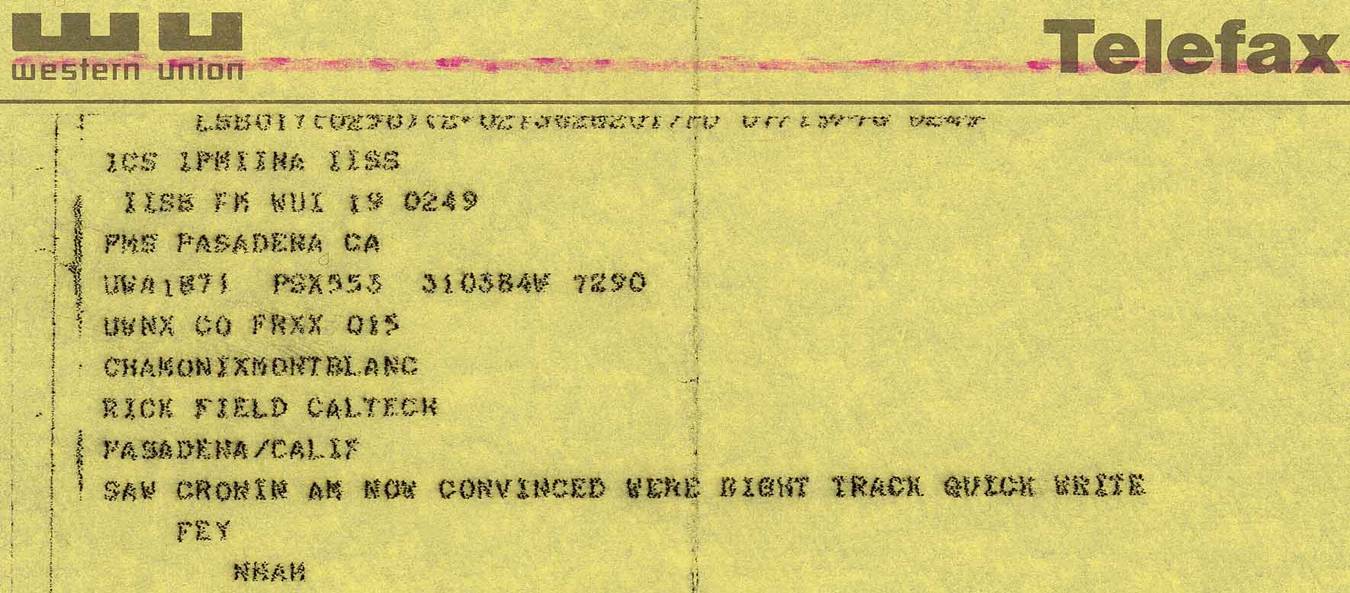}
\caption{ A July 1976 telegram from Feynman in France to Rick Field at
  CalTech, urging Rick to write
  up their work on parton-parton scattering followed by parton
  fragmentation being the source of the high-Pt particles (Courtesy of
  R. Field).}
\label{fig:feynman_to_field}
\end{figure}

However the interpretation of the `excess' and energy-dependence of
high-\Pt particle production as two-body parton scattering
was not yet agreed on. Figure~\ref{fig:feynman_to_field} shows a
telegram from Feynman to Rick Field after Feynman and Cronin talked
about the high \PT particle production results in France. The telegram
is excited, frankly, about the observation of parton scattering, with
its consequent $1/P_T^4$ dependence, followed by parton fragmentation
to give the observed $1/P_T^8$ at Fermilab.

\section{The Chaotic Start of a Collider Program at Fermilab}

\begin{figure}[h]
\centering
\includegraphics[angle=0,width=0.85\textwidth]{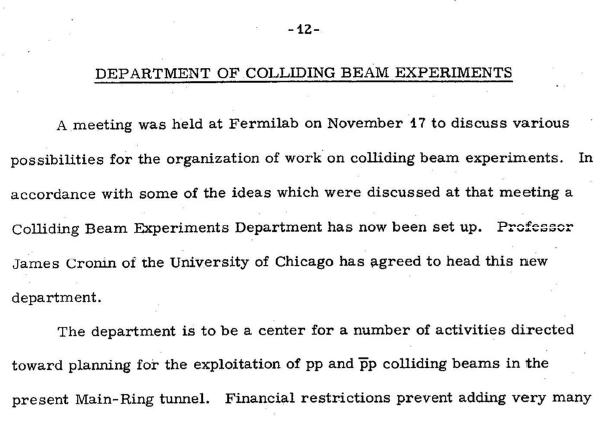}
\caption{A memo announcing the start of the Fermilab Colliding Beam
  Department, with Jim Cronin as Head.}
\label{fig:colliding_dept_start}
\end{figure}

With the initiatives of Rubbia and others at CERN the pressure grew on
Fermilab toward a collider. Bob Wilson appointed Jim Cronin as Head of
a new department, the Fermilab Colliding Beam Department. Cronin
started a design for the detector, and started some basic measurements
of particle flux in the MR tunnel.
Figure~\ref{fig:colliding_dept_start} is from the memo announcing the
start of the Department.

\begin{figure}[h]
\centering
\includegraphics[angle=0,height=0.60\textheight]{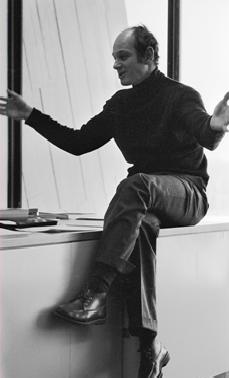}
\caption{Jim Cronin, the  Head of the new Fermilab Colliding Beam
  Department. Cronin started the first designs for the detector and
  measurements in the MR tunnel.}
\label{fig:jwc}
\end{figure}

\begin{figure}[h]
\centering
\includegraphics[angle=0,width=0.98\textwidth]{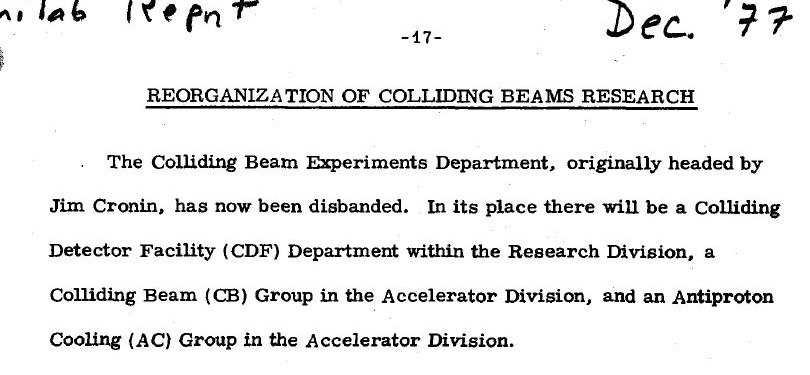}
\caption{A memo announcing the end of the Fermilab Colliding Beam
  Department. The path to a realistic collider with an adequate
  luminosity, built on a credible schedule,  and a
sophisticated detector adequate for the remarkable physics potential
was not clear, and Cronin had resigned. }
\label{fig:colliding_dept_end}
\end{figure}

Unfortunately a rational approach didn't work at Fermilab, partly
because of Federal HEP funding priorities, partly because of
personnel, and partly because of organizational structure.
Figure~\ref{fig:colliding_dept_end} shows the announcement of
Cronin's resignation.

Some background is in order to understand the chaotic situation and
the impact the international collaborators from Pisa/Frascati and Japan had in
fashioning a successful program out of the initial conditions. The
Main Ring (MR) accelerator was not well-suited as a collider. With
remarkable foresight Wilson had made provisions for the
superconducting Tevatron ring in the tunnel, mounted below the
MR. However, there were a number of competing facility proposals: PEP
at SLAC, Isabelle at BNL, and CESR at Cornell. The HEPAP panel held
Sub-panel meetings at Woods Hole in 1974 (Weisskopf, Chair), 1975
(Low, Chair), and 1977 (Sandweiss, Chair) to evaluate and prioritize
the proposals. The panels proved that they could differentiate,
although with the sign wrong: we approved PEP and Isabelle for
construction, leaving the (successful!) Tevatron and CESR to be funded by
other means.

\begin{figure}[h]
\centering
\includegraphics[angle=0,width=0.95\textwidth]{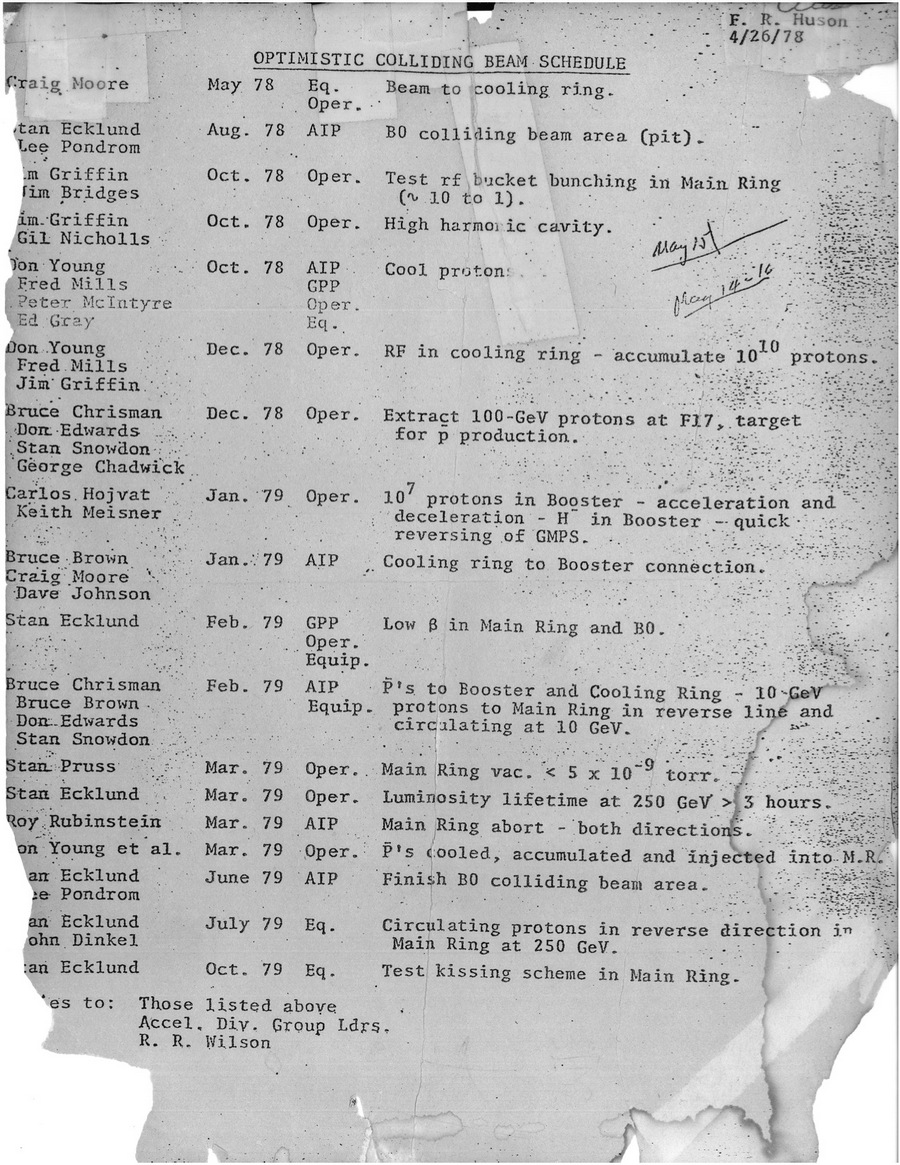}
\caption{An April 1978 ``Optimistic Colliding Beam Schedule'' for
  \pbarp collisions in the Main Ring, with first collisions planned
  for 18 months later. This was optimistic in many ways. }
\label{fig:optimistic_schedule}
\end{figure}

\clearpage

\begin{figure}[h]
\centering
\includegraphics[angle=0,height=0.55\textheight]{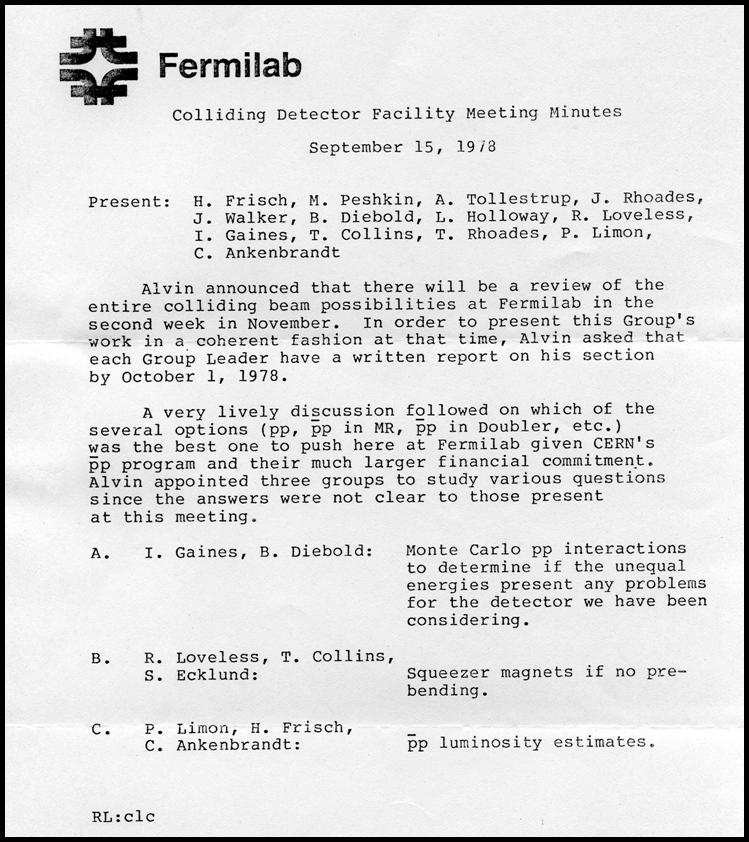}
\caption{The minutes of a 1978 CDF meeting
  discussing the options for the accelerator itself (``pp, \pbarp in
  MR, \pbarp in Doubler, etc.''. The words ``A very lively discussion
  followed'' are a diplomatic way of saying that it was a mess, with
  many competing options and unsupported claims of luminosities,
  schedules, and costs. Note the last sentence and the three groups.}
\label{fig:1978_minutes}
\end{figure}

Figure~\ref{fig:1978_minutes} illuminates the lack of consensus on
what to do. Among the options for the accelerator were colliding the
Main Ring with the Argonne ZGS (yes, really) for proton-proton collisions. The
estimates of achievable luminosities for several of the options
disagreed by orders-of-magnitude. The options for the detector were
equally all over the map; one proposal had a detector in a pit barely
big enough to hold it, but on cables so that it could be raised for
access.

\clearpage


\begin{figure}[h]
\centering
\includegraphics[angle=0,width=0.70\textwidth]{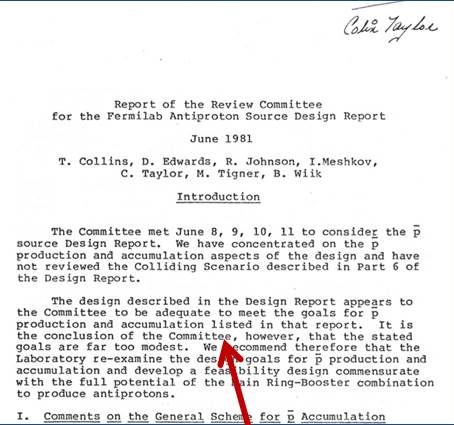}\\
\vspace*{0.75in}
\includegraphics[angle=0,width=0.65\textwidth]{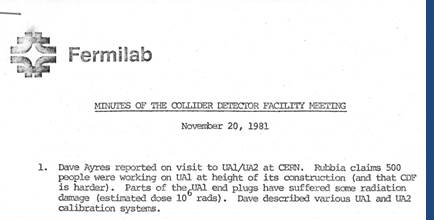}
\caption{Top: The June report of the 1981 distinguished review
  committee on the Fermilab Antiproton Source. Bottom: An excerpt from
  the minutes CDF meeting the following November, showing the
  awareness at Fermilab of the rapidly-moving CERN program. Both the
  UA1 and UA2 trigger and tracking designs had impact on the CDF
  designs; we studied them and wanted to do better.}
\label{fig:acc_options}
\end{figure}

\clearpage

\begin{figure}[h]
\centering
\includegraphics[angle=0,width=0.80\textwidth]{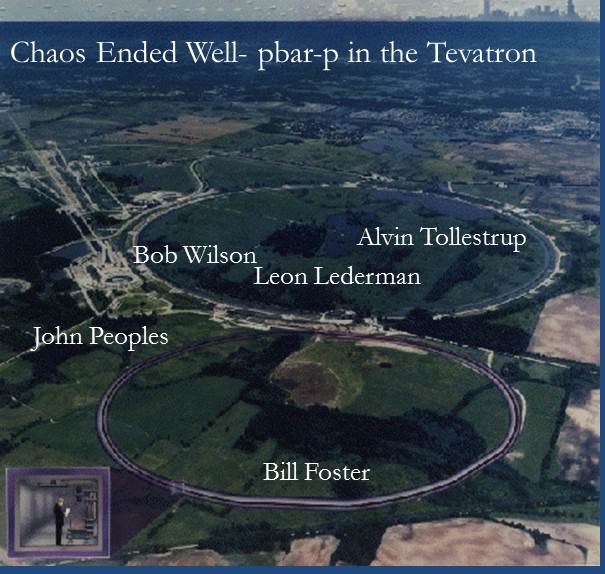}
\caption{The completed Fermilab accelerator complex showing the
  remarkable solutions by Wilson, Tollestrup, Peoples, and many others
  to the lack of adequate funding, the technical
  challenge of a large ring high-field superconducting magnets, the
  pbar source, and the development of a sophisticated hadron
  collider detector with both precise tracking and projective calorimetry. }
\label{fig:chaos_ended_well}
\end{figure}

That a technically sound and exceptionally successful solution came
out of the chaos was due to a number of remarkable people
and the very strong staff and attitude at the Lab.
Key elements were Wilson's plans for a
superconducting ring in the MR tunnel, Alvin Tollestrup's solutions to
the superconducting magnet problems, and John People's leadership of the
pbar source. Figure~\ref{fig:chaos_ended_well} shows the current
Fermilab  facility, including the Recycler storage ring  for
which Bill Foster played an essential role.

\clearpage

\section{The Foundational Contributions of Pisa to the CDF Detector Design}

In his contribution to this conference, Giorgio Bellettini  describes
how in 1979 he and Paolo Giormini met with Alvin Tollestrup and Bob
Diebold to discuss Pisa and Frascati joining CDF.

\begin{figure}[h]
\centering
\includegraphics[angle=0,height=0.40\textheight]{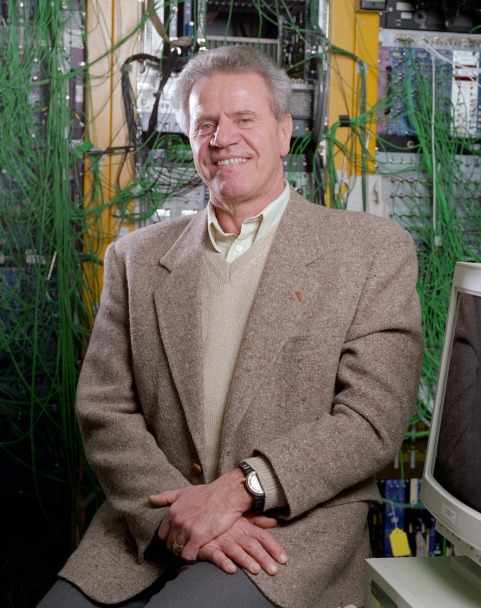}
\caption{Giorgio Bellettini, the leader of the Pisa CDF group, in the
  CDF Trigger Room in 1995. One of the Silicon Vertex Trigger (SVT)
  racks is visible behind him.}
\label{fig:giorgio}
\end{figure}

\begin{figure}[h]
\centering
\includegraphics[angle=0,width=0.98\textwidth]{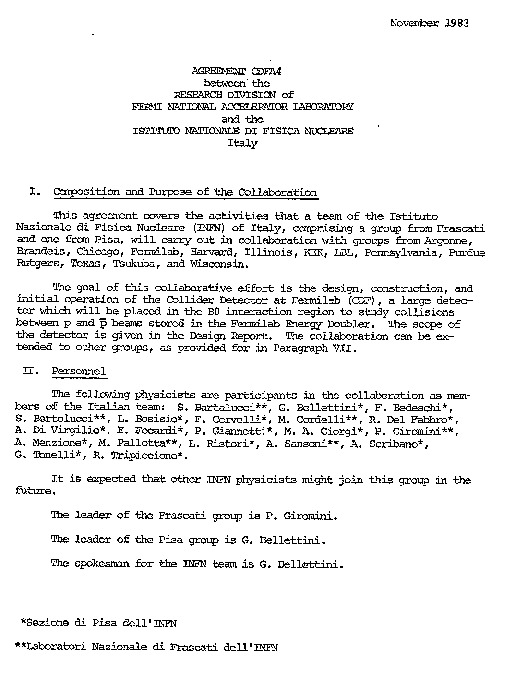}
\caption{The 1983 Memorandum of
Understanding (MOU) between Pisa/Frascati and Fermilab. The group was
both large and very strong, bringing collider experience and
extraordinary detector development skills to CDF.}
\label{fig:1983_MOU}
\end{figure}

Figure~\ref{fig:1983_MOU} is a copy of the 1983 Memorandum of
Understanding (MOU) between Pisa, Frascati, and Fermilab. Each of the
collaborating institutions ended up signing such an MOU. Note the
original cast members. The original institutions included Tsukuba and
KEK in Japan, with the strong Japanese effort led by Kuni Kondo.

\begin{figure}[h]
\centering
\includegraphics[angle=0,width=0.98\textwidth]{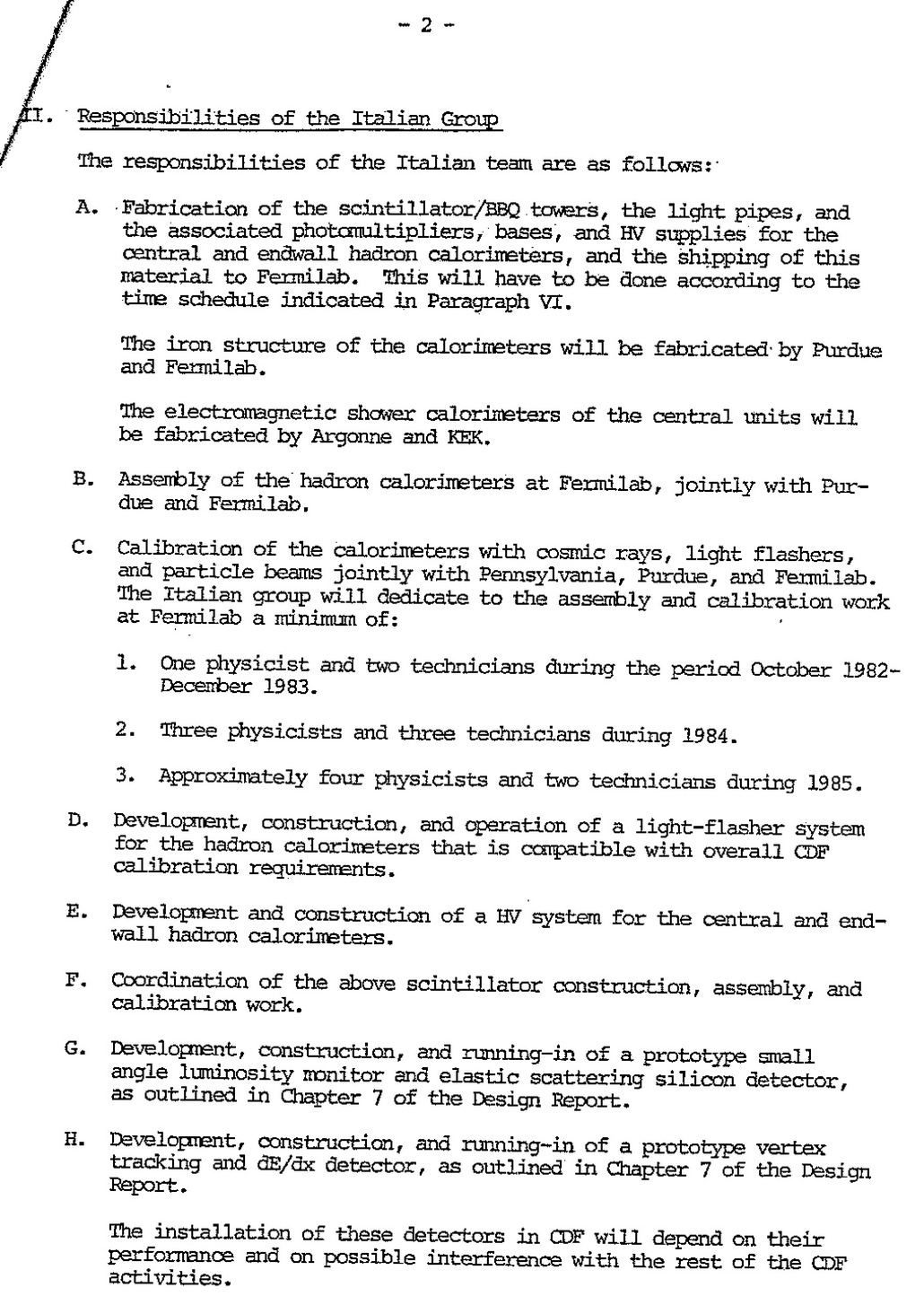}
\caption{The list from the 1983 MOU of Pisa and Frascatti
  responsibilities.}
\label{fig:1983_MOU_list_of_responsibilities}
\end{figure}

Figure~\ref{fig:1983_MOU_list_of_responsibilities} shows the essential
role played by Pisa and Frascatti in the design, construction, and
subsequent operation of CDF by listing the extensive responsibilities
of the group. The remarkable characteristics of the tracking and
calorimeter systems were heavily influenced (or more) by the group.

\clearpage

\subsection{The Technical Design Report (TDR)}
\begin{figure}[h]
\centering
\includegraphics[angle=0,width=0.42\textwidth]{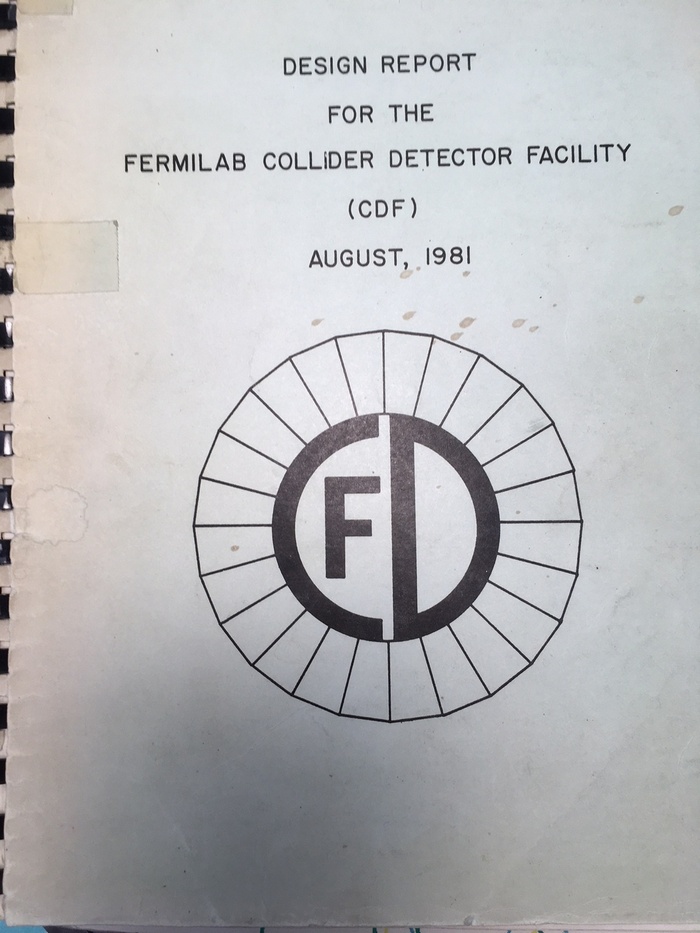}
\hfil
\includegraphics[angle=0,width=0.50\textwidth]{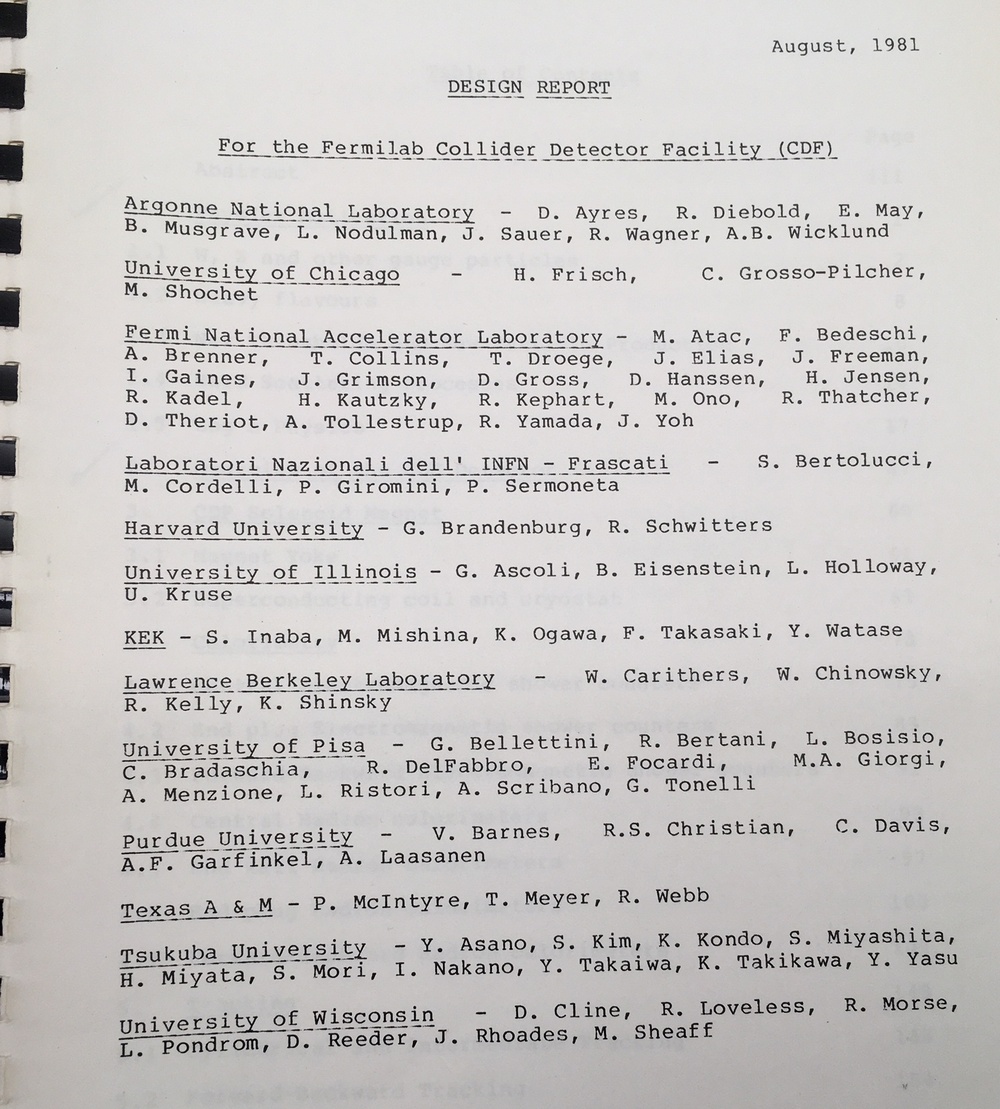}
\caption{The cover and author list of the Aug. 1981  Technical
Design Report.} \label{fig:TDR_cover_authors}
\end{figure}

Design and even construction of the Collider Detector Facility, as it
was initially named, started before a proposal had been
written~\footnote{When a group of us, probably led by Tollestrup,
  presented the idea to Bob Wilson, he said ``OK, but it can't cost
  more than 10 million dollars''. We knew, and I'm sure he knew, that
  this was low by an order-of-magnitude; we looked at each other, and
  said ``sure''.}. Instead, the Collaboration wrote a Technical Design
Report (TDR), released in August of 1981~\cite{CDF_TDR}.
Figure~\ref{fig:TDR_cover_authors} shows the cover, with its logo
of the projective calorimeter towers put forward by the Pisa
group.

\begin{figure}[h]
\centering
\includegraphics[angle=0,width=0.98\textwidth]{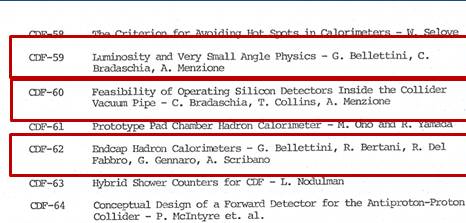}
\caption{Key 1981 CDF Notes (the internal collaboration technical
  papers) The CDF Note index showing the 1981 CDF Notes from the Pisa/Frascatti groups,
  proposing designs for silicon detectors and hadron calorimeters. The
  titles of notes CDF-59, CDF-60, and CDF-62 are highlighted in
  red.}
\label{fig:pisa_cdf_notes}
\end{figure}

The titles of the 1981 CDF Notes (internal collaboration technical
notes) proposing designs for the vertex silicon detectors and
hadron calorimeters are shown in Figure~\ref{fig:pisa_cdf_notes}.
The as-built detectors corresponded remarkably faithfully to these
initial innovative designs.

\subsection{Projective Hadron Calorimetry: Pisa and Frascatti}

\begin{figure}[h]
\centering
\includegraphics[angle=0,width=0.48\textwidth]{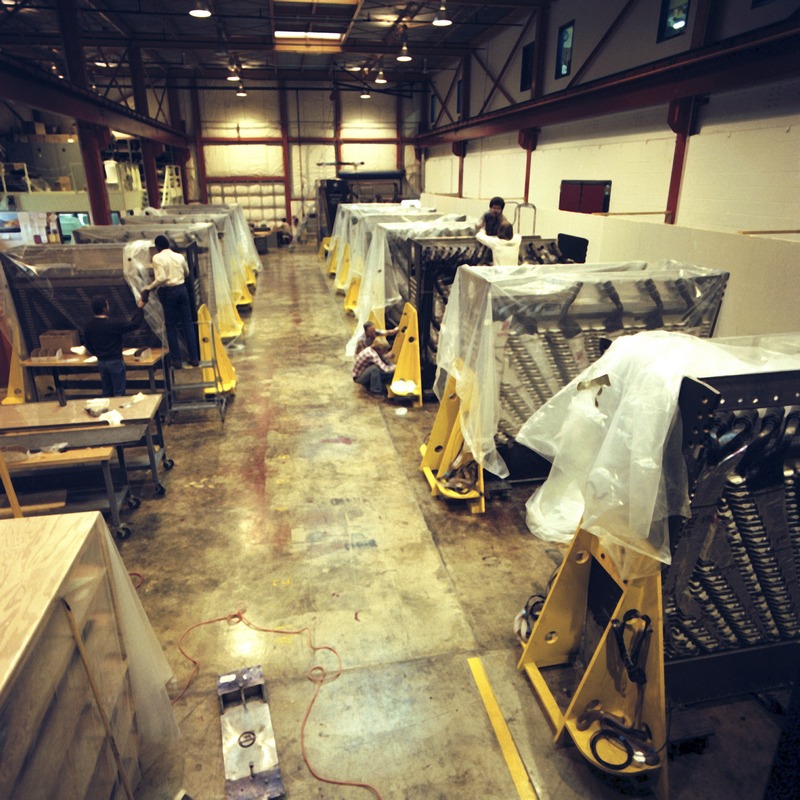}
\hfil
\includegraphics[angle=0,width=0.48\textwidth]{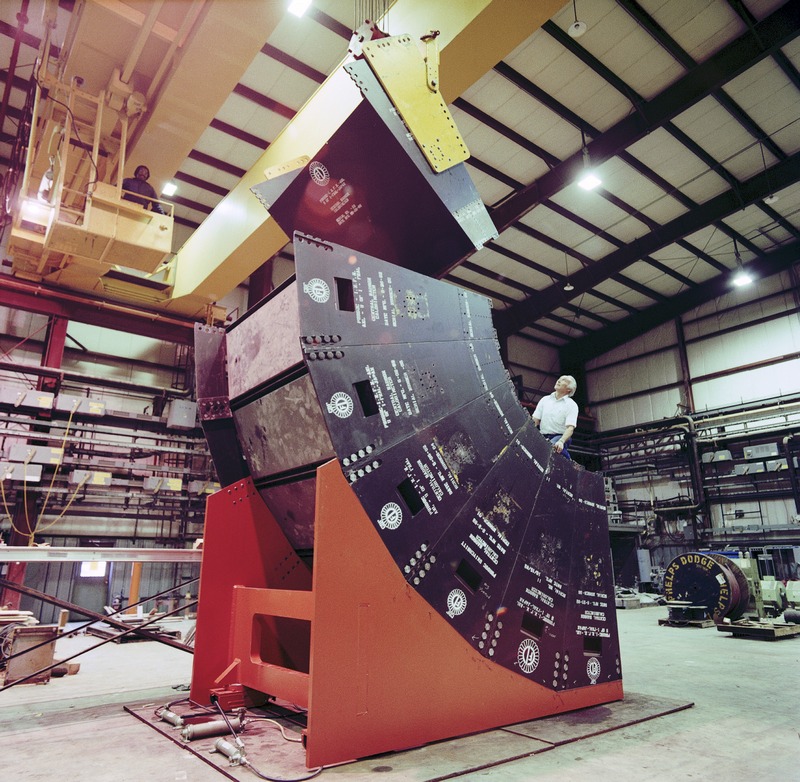}
\caption{Left: Franco Cervelli and Aldo Menzione working on hadron
  calorimeter modules. Right: Hans Kautsky supervising the assembly of
  prototype `arch' of modules being assembled at Argonne. (Note the hard-hat).}
\label{fig:Pisa_Frascati_HCAL}
\end{figure}

A very strong contingent from Pisa and Frascatti, in collaboration
with Argonne, Fermilab, Purdue, Urbana, and other institutions, led
the adoption and design of the (largely) hermetic calorimeters based
on a projective tower geometry of uniform segmentation in rapidity and
polar angle. Transporting the light out of the calorimeter to the
photomultipliers outside was a tour-de-force, both in design and
implementation.  The left-hand panel of
Figure~\ref{fig:Pisa_Frascati_HCAL} shows Franco Cervelli and Aldo
Menzione working on hadron calorimeter modules. The right-hand panel
shows a prototype of one of the four `arches' of modules being
assembled at Argonne~\footnote{No mention of the detector construction
  would be complete without due credit being given to Dennis Theriot as
  overall CDF Project Manager, and Hans Jensen leading the detector
  assembly in the $B0$ collision hall.}.

\subsection{Silicon Vertex Detectors (SVX) and Silicon Vertex Trigger (SVT)}

\begin{figure}[h]
\centering
\includegraphics[angle=0,width=0.98\textwidth]{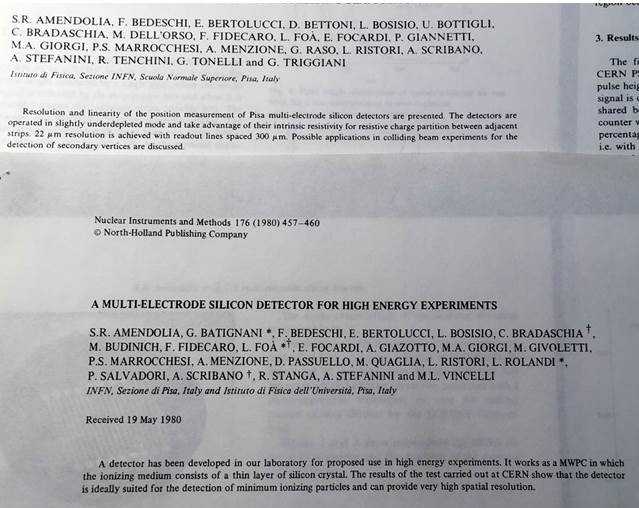}
\caption{Two seminal papers by the Pisa group on the exploitation of silicon vertex detectors.}
\label{fig:Si_HCAL}
\end{figure}

Aldo Menzione was a key pioneer in the development of silicon
tracking detectors, as well as single-handedly, it seemed, bulling
through the adoption by CDF of the SVX silicon detector that
played such an important role in the Tevatron physics program.
Figure~\ref{fig:Si_HCAL} reproduces title from papers in the
development

\clearpage

I was on the tracking `GodParent Committee', led by Giulio Ascoli and
later Willie Chinowsky, that had the charge of vetting the tracking
designs.  The history of vertex tracking at CERN had not been good,
and there was nervousness, particularly among the less technical folk;
Aldo was an irresistible force, and the GodParents approved the
installation of the silicon vertex detector. The SVX, along with the
remarkable central tracking chamber of Kadel, Mukherjee, Kephart,
Binkley, and others, were the core of CDF performance.

\begin{figure}[h]
\centering
\includegraphics[angle=0,width=0.98\textwidth]{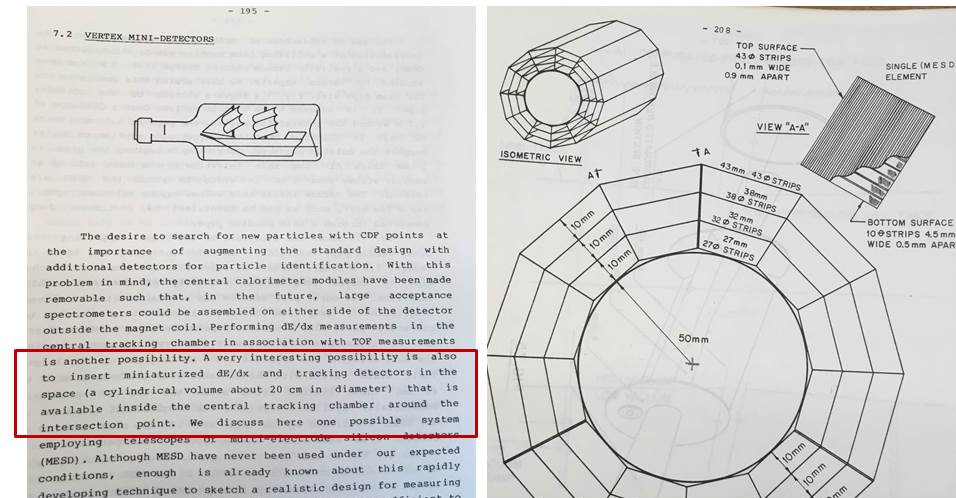}
\caption{The description and layout of the Silicon Vertex Detector (SVX) in the
  Technical Design Report.}
\label{fig:SVX_TDR}
\end{figure}

Aldo's design and description in the Technical Design Report
of this bold proposal are shown in Figure~\ref{fig:SVX_TDR}. What is
truly remarkable to me is how this sketch corresponds to what was
built and operated; even though this was completely unexplored
territory for us, Aldo knew deeply what the system needed to do.

\clearpage

\begin{figure}[h]
\centering
\includegraphics[angle=0,width=0.48\textwidth]{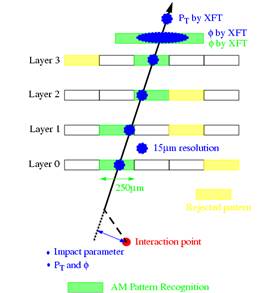}
\hfil
\includegraphics[angle=0,width=0.48\textwidth]{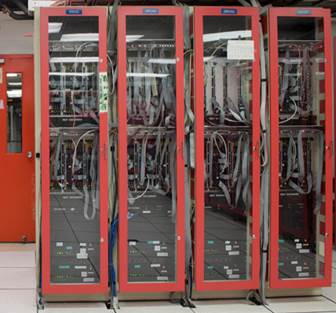}
\caption{Left: The operating principle of the Silicon Vertex Tracker
  (SVT). Right: the SVT system installed in the CDF Trigger Room.}
\label{fig:SVT_Luciano}
\end{figure}

The ability to trigger on displaced vertices, seemingly impossible
when we first contemplated the collider, was brilliantly implemented by
Luciano Ristori and the Pisa group (Chicago collaborated on the
hardware). The precision, data rate, and necessary monitoring/control
were a large step beyond the
state-of-the-art. Figure~\ref{fig:SVT_Luciano} (Left) shows a
`cartoon' of the operating principles; the installation, driven by
over 100 optical fibers running at 1.4 Gbits/sec, is shown on the right.

\subsection{Summary}

The period discussed here spans the development of precision
measurements at hadron colliders and the parton model.
The Pisa/Frascatti group's contributions to CDF have set the standards for
collider programs at the `Energy Frontier': the precision spectrometer
consisting of precision tracking back to the vertex with a
silicon vertex detector followed by projective segmented calorimeters,
and using a trigger on displaced vertices.  The combined tracking and
calorimeter systems that the group contributed so much to allowed
precision calibration of both systems through the `E/p' method, now a
standard (as are so many of the innovations from CDF) at the LHC.

\begin{figure}[h]
\centering
\includegraphics[angle=0,width=0.48\textwidth]{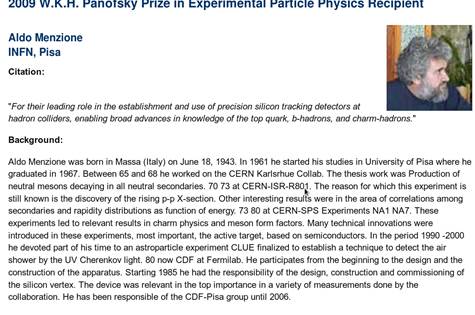}
\hfil
\includegraphics[angle=0,width=0.48\textwidth]{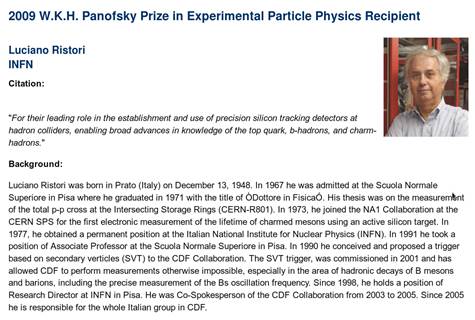}
\caption{The Panofsky Prize citations for Aldo Menzione and Luciano Ristori.}
\label{fig:Panofsky_prizes}
\end{figure}

The work of Aldo Menzione on the SVX and Luciano Ristori were honored
with the Panofsky Prize of the American Physical Society.
Figure~\ref{fig:Panofsky_prizes} gives the citations. Recognition of
the importance of the entire group has been recognized by the
continued election of Pisa physicists to the leadership
of the Collaboration.

In summary, this was a remarkable group of wonderful physicists and people. It
was an honor and pleasure to work closely with them; the leadership
and exceptionally high  intellectual standard was key to the successes
of CDF.




\section{Acknowledgements}
I would like to thank the Organizers for the enjoyable opportunity
to revisit wonderful memories of collaboration with the Pisa CDF
physicists. I owe deep thanks to Vincenzo Cavasinni for his care
and thoughtful hospitality, and Giorgio Bellettini for the
invitation and help. The occasion to see old friends from CDF was
particularly special. The talk is dedicated to the memories of
Aldo Menzione and Mauro Dell'Orso. I also want to honor Kuni
Kondo, who headed the Japanese group which joined CDF at the same
time the Pisa/Frascatti group.


\end{document}